# Suppression of Magnetic Phase Separation in Epitaxial SrCoO$_X$ Films

F. J. Rueckert[1], Y. F. Nie[1], C. Abughayada[2], S. A. Sabok-Sayr[2], H. Mohottala[3], J. I. Budnick[1], W. A. Hines[1], B. Dabrowski[2], and B. O. Wells[1]

[1] Department of Physics, University of Connecticut, Storrs, Connecticut 06269, USA
[2] Department of Physics, Northern Illinois University, DeKalb, Illinois 60115, USA
[3] Physics Department, University of Hartford, West Hartford, Connecticut 06117, USA

Using pulsed laser deposition and a unique fast quenching method, we have prepared SrCoO$_x$ epitaxial films on SiTiO$_3$ substrates. As electrochemical oxidation increases the oxygen content from x = 2.75 to 3.0, the films tend to favor the discrete magnetic phases seen in bulk samples for the homologous series SrCoO$_{(3-n/8)}$ (n = 0, 1, 2). Unlike bulk samples, 200nm thick films remain single phase throughout the oxidation cycle. 300 nm films can show two simultaneous phases during deoxidation. These results are attributed to finite thickness effects and imply the formation of ordered regions larger than approximately 300 nm.

Many transition metal oxides exhibit a rich phase diagram as electron interactions result in multiple ground states with comparable energy. Of particular interest are doped correlated-insulators, where charge doping, often leads to electronic inhomogeneity and in some cases results in clear electronic phase separation.[1] An important example is the perovskite La$_{1-x}$Sr$_x$CoO$_3$ (LSCO), in which the substitution of Sr$^{2+}$ for La$^{3+}$ leads to a magnetically inhomogenous state and an insulator to metal transition.[2–4] An alternative method for controlling the cobalt valence is through the amount of negatively charged oxygen present, altering the balance between ground state energies and modifying the crystal structure. In this case the insulating, parent compound is SrCoO$_{2.5}$ with oxygen concentration controllable up to SrCoO$_3$.

Recently, magnetic phase separation has been reported in bulk samples of SrCoO$_x$ for oxygen content 2.88 < x < 3.0.[5] Beginning with x = 2.88, samples were electrochemically oxidized to x = 3.0, while magnetic measurements were made at stages along the process. At the endpoints, the x = 2.88 and x = 3.0 compounds show single ferromagnetic (FM) transitions at T$_C$ = 220 K and 280 K, respectively. At intermediate concentrations, both magnetic transitions are seen, while high resolution x-ray diffraction (XRD) shows only a single crystallographic phase. This implies an electronic mechanism of separation, where the stable phases may be related to an ordering of either oxygen vacancies or Co ion valence. In this work, we describe the suppression of phase separation in films below a critical thickness of 200 nm. The films tend to stabilize with ferromagnetic transitions at similar temperatures as in the bulk, indicating that the same phases are most stable. Taken together, this implies phase separation is related to the formation of distinct magnetic regions of large dimension.

The evolution of the SrCoO$_x$ crystal structure with oxygenation has been detailed in neutron diffraction studies.[6] SrCoO$_{2.5}$ is orthorhombic, with the brownmillerite-type structure, and orders antiferromagnetically at T$_N$ = 570 K.[7] As oxygen is incorporated, a second cubic structure develops, corresponding to SrCoO$_{2.75}$, which has been identified as a ferromagnet with ordering temperature T$_C$ = 160 K.[5] This phase grows at the expense of the orthorhombic phase until the sample is completely SrCoO$_{2.75}$.[8] Above x=2.75, the material remains cubic and the lattice constant evolves smoothly, with the exception of a slight tetragonal distortion near x=2.88.[6,8] At x=3.0, SrCoO$_x$ is a ferromagnet with T$_C$=280K.[9]

Single crystalline samples can provide more detailed information concerning crystalline orientation and mitigate the effects of grain boundaries. While such crystals have been prepared,[10] the growth process makes analysis with oxygen evolution difficult. Epitaxial films offer a practical alternative. One report describes the growth of SrCoO$_{2.5}$ films which can be fully oxidized through a wet chemical process.[11] We have used a novel post growth anneal and quenching technique to produce metastable films of SrCoO$_{2.75}$. Electrochemical oxidation is used to increase oxygen content up to x = 3.0. In an effort to identify the effects of sample ordering and dimensionality in this system, we preformed magnetic measurements on epitaxial thin films of SrCoO$_x$ grown on SrTiO$_3$ (0 0 1) and (1 1 1) substrates.

Films were grown using pulsed laser deposition (PLD) from a SrCoO$_{2.5}$ target. Dense polycrystalline targets were prepared by solid state synthesis from SrCO$_3$ and Co$_3$O$_4$. Pressed powders were fired several times in air at increasing temperatures from 900 to 1150 °C, followed by intermittent grinding. Final firing of a d = 24 mm target was done as follows: 24 h at 1150 °C, slowly (1 deg/min) cooled to 900 °C, followed by quenching to room temperature on a copper plate. The resulting material was single-phase Brownmillerite with oxygen content near 2.5.

A 248 nm KrF excimer laser operated at 4 Hz was used to deposit the films on SrTiO$_3$ (STO) substrates of (0 0 1) and (1 1 1) orientation. The substrates were held at 700 °C in an atmosphere of 300 mTorr O$_2$ and cooled slowly after growth in 300 Torr O$_2$. To achieve epitaxial films of single phase, a secondary annealing and quenching process was required. The films were annealed in the PLD chamber for 30 min at 750 °C in 300 Torr $O_2$, then quickly cooled in vacuum to ambient temperature through contact with an internal, water cooled stage. This process resulted in well-ordered films of nominal oxygen concentration x = 2.75, with $T_C$ = 160K. Electrochemical oxidation was performed using a home built potentiostat and a three wire electrochemical cell. The electrodes were submerged in an alkali electrolyte solution of 0.05 M NaOH and held at a constant voltage difference of 0.8 V. We produced a number of film samples with oxygen $2.75 \leq x \leq 3.0$. Structural characterization was carried out using conventional Cu Kα XRD using a Siemens D8 θ-2θ powder diffractometer. Magnetic ordering temperatures were measured at various oxidation intervals using a Quantum Design MPMS SQUID.

Figure 1 shows typical XRD patterns for films used in our oxidation studies. The films are oriented in the (0 0 1) (panel a) and (1 1 1) direction (panel b), with no other orientations evident. Though partially obscured by the much stronger substrate peaks, the observed film peaks are indexed as a perovskite pseudo-cubic lattice. Using the known $SrTiO_3$ substrate as a reference (a = 3.905 Å), the film peaks for both films fit a lattice constant of a = 3.83 Å, in agreement with the bulk and previous reports,[6,8–10,12] indicating the film is fully relaxed.

Film thickness is estimated through an analysis of magnetic behavior as a function of applied field (not shown). The data approach linear behavior with negative slope as the applied field is increased. Extrapolation from the liner regions back to the vertical axis (H = 0) yield values for the film saturation magnetic moment, while the slopes give the substrate diamagnetic susceptibility. Accounting for the substrate, an average value of $2.0 \times 10^{-3}$ emu is measured for the saturation magnetic moment due to the film. Using the measured value of a = 3.83 Å, the film density was estimated to be ρ = 5.7 g/cm$^3$. Assuming that the saturation magnetization of the film is the same as that for bulk $SrCoO_3$ (σ = 63.5 emu/g or 2.2 $\mu_B$/Co)[13], film thicknesses of 200 nm and 300 nm (± 20 nm) are estimated for the samples described in this study. These estimates are consistent with spot checks via scanning electron microscopy of film profiles.

The relatively large deposition thickness of greater than 200 nm and concurrence with the bulk lattice constant suggest substrate induced strain is not significant in these samples. We could achieve full oxidation (x=3.0) after approximately 20 minutes at 0.8 V, in contrast to the 24 hours required for bulk samples. Oxygen content is estimated by comparing the sample ordering temperature to the observed $T_C$ of the bulk,[5] correlated with the electrochemical literature[6,8] by XRD measurements of lattice constant. Magnetization data shows ordering temperatures for the as grown samples are $T_C$ = 160 K (1 0 0) and 180 K (1 1 1), indicating an oxygen content of approximately x = 2.75 when compared to the literature.[5]

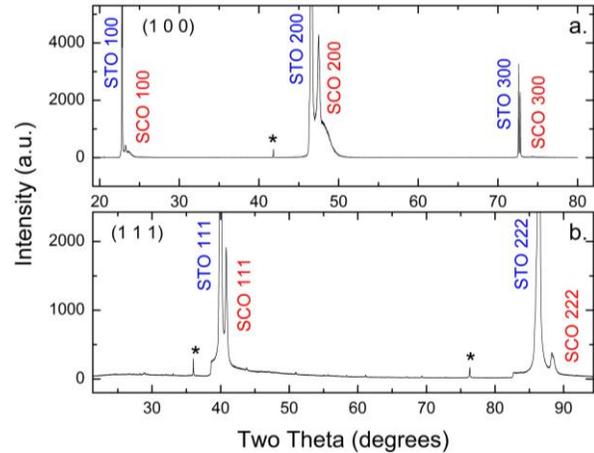

Figure 1 X-ray diffraction pattern of (a) (1 0 0) oriented 200 nm film and (b) (1 1 1) oriented 300 nm film samples used in oxidation study. (*) Represents Cu $K_\beta$ radiation. Using the STO substrate peaks as a reference, the observed SCO film peaks correspond to a pseudo-cubic lattice constant of a = 3.83 Å. No other phases are observed.

Samples were electrochemically oxidized in stages up to a total exposure time of 20 minutes. To track possible changes in magnetic phase as the electrochemical process progressed, magnetization versus temperature data were taken at stages up to full oxidation. Figure 2a shows field cooled data at 3, 5, 10, 15, and 20 min increments for the (1 0 0) film. Stable magnetic phases are notable at $T_C$ = 160 K, 230 K, and 270 K, roughly matching the phases described by Xie *et al.*.[5] However, unlike in the bulk, no two-phase behavior is seen during oxidation. Figure 2b shows comparable phases for a set of (1 1 1) ordered films. Transition temperatures of $T_C$ = 180 K, 210 K, and 270 K are measured for as grown, intermediate, and fully oxidized samples, respectively, also in close agreement with the noted bulk phases. The variation in transition temperatures indicates a small range over which each phase is stable.

Beginning with the as grown films, oxidation raises the moment, with a significant increase in magnitude at full oxidation. Though the transition temperature also increases, the coexistence of two magnetic phases was not observed for films of any orientation or thickness. Rather than the growth of one phase at the expense of the previous, as reported for the bulk,[5] the film samples tend to remain at particular stable phases until, with the addition of enough oxygen, the sample transitions to the next $T_C$ phase. As seen in Figure 2a, although the data for oxidation of 10, 15, and 20 min were taken at equal increments, a transition to a higher ordering temperature only occurred after the final stage. While the oxygen absorbed between 10 and 15 minutes was insufficient to effect a change in $T_C$, an equal increment of further exposure results in the shift noted at 20 minutes. Thus oxygen absorption occurs throughout the process, but the transition to the next phase is dependent on the necessary electronic balance.

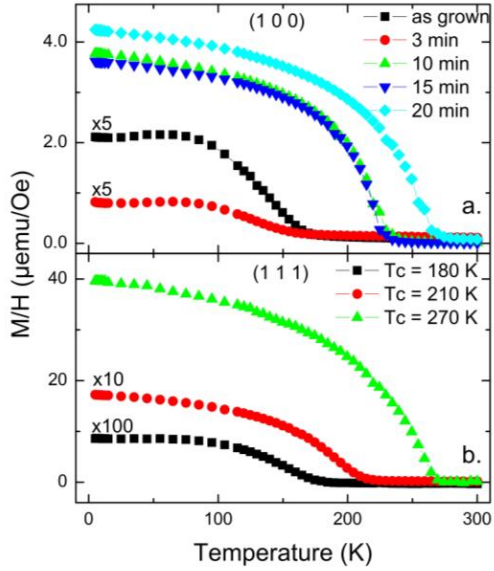
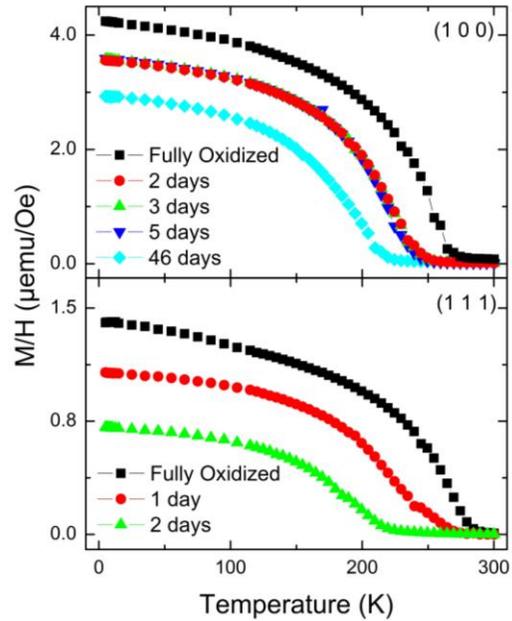

Figure 2 (a) Field cooled magnetization vs temperature showing $T_C$ as a function of oxidation for a single (1 0 0) film after successive oxidation steps. Three distinct transitions are evident, at $T_C$ = 160 K, 230 K, and 270 K. Times describe the cumulative number of minutes the sample was electrochemically oxidized. (b) Field cooled magnetization of (1 1 1) oriented films at different stages of oxidation. Similar phases are noted at $T_C$ = 180 K, 210 K, and 270 K. No two-phase behavior is seen for either orientation. Curves are scaled as indicated for clarity.

Figure 3 Field cooled magnetization vs temperature as oxygen leaves the films after various times since growth for (a) (1 0 0) and (b) (1 1 1) oriented films of 200 nm thickness. Times indicate the total number of days after full oxidation. (a) The x = 3.0 phase ($T_C$ = 270 K) decays quickly to the more stable x = 2.75 ($T_C$ = 240 K) phase. This phase degrades much more slowly to one with lower $T_C$. (b) A (1 1 1) oriented film of the same thickness transitions from fully oxidized $T_C$ = 270 K to $T_C$ = 220 K after 2 days in air. Throughout the process, all films of 200 nm thickness remain single phase.

$SrCoO_3$ is oxidized well past equilibrium and not stable in ambient conditions. Thus we can also follow the magnetic phase behavior during natural deoxidation. This process is dependent on the thickness of the film. The stability of the fully oxidized phase for (1 0 0) and (1 1 1) films of 200 nm thickness is shown in Figure 3. In contrast to the bulk, the x = 3.0 phase loses oxygen relatively quickly. For the (1 0 0) film (Fig 3a), after two days the sample is no longer fully oxidized, and decays to the more stable $T_C$ = 220 K (x = 2.88) phase. Similarly, the (1 1 1) film (Fig 3b) transitions to $T_C$ = 210 K over the same time frame. For both orientations, a two-phase state was not observed after deoxidation.

Figure 4 shows the magnetic state of a fully oxidized (1 1 1) film of 300 nm thickness after 15 days of exposure to ambient atmosphere. Unlike the 200 nm case, two coexisting phases are observed, with $T_C$ = 175 K and 285 K. The inset of Figure 4 reproduces the phase diagram from Ref [5]. It appears that, for films of this thickness, the x = 2.88, $T_C$ = 220 K phase is less likely to appear than those with higher or lower $T_C$. Nevertheless, it is the first sign of the reappearance of the two phase magnetic behavior that characterizes the bulk.

Generally, there can be several differences between bulk samples and well-ordered epitaxial films. The most commonly discussed film-specific issues are finite size effects on magnetism or strain caused by lattice mismatch with the substrate. For similar $La_{0.5}Sr_{0.5}CoO_3$ films these effects are typically significant for films thinner than about 60 nm.[14] A kind of phase separation has been reported in the near-interface region of some LSCO films, though the effect is only for regions on the order of 10 nm.[15] Thus, both the 200 and 300 nm films used in this study should be considered as bulk-like in most respects, but with a limited size in one dimension.

The symmetry of the cubic structure and lack of significant strain suggest the suppression of phase separation for 200 nm films is also not due to classic film effects. The suppression is observed for films of both (1 0 0) and (1 1 1) alignment, making orientation an unlikely cause. Instead, we conclude the ground state energy of the magnetic phases is affected by thickness. At 200 nm, the films tend toward a metastable $T_C$ = 240 K phase, whereas the 300 nm films stabilize at the $T_C$ = 175 K and 285 K end points. Therefore, the two phases seen in the 300 nm thick film indicate a size dependent effect with a critical thickness of approximately 300 nm.

The coexistence of magnetically distinct phases within a single chemical composition has been described in the related material $La_{1-x}Sr_xCoO_3$.[2,16] In that system, hole-rich FM clusters form within a hole-poor matrix, with correlation lengths on the order of 10 nm.[3,17] It may be that the cause of phase separation is similar but that the larger length scale noted in this work is a result of the type of dopant introduced, i.e., mobile oxygen ions rather than substitutional cations.[18]

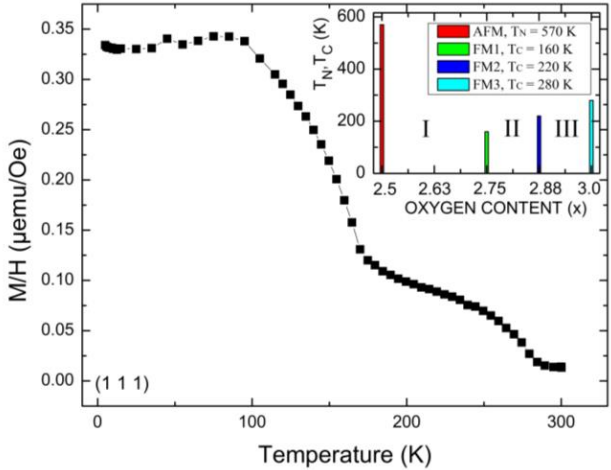

Figure 4 Temperature dependent magnetization of a fully oxidized, 300 nm thick (1 1 1) oriented film. Data was taken 15 days after oxidation. As in the bulk, two magnetic phases are evident with $T_C$ = 175 K and $T_C$ = 285 K. These correspond roughly to oxygen concentrations of x = 2.75 and x = 3.0. The inset shows the phase diagram described for bulk samples, reproduced from Ref 5.

$SrCoO_x$ and $La_{1-x}Sr_xCoO_3$ form a system with similarities to the cuprate superconductors $LaCuO_{4+y}$ and $La_{2-x}Sr_xCuO_4$. For the cuprates, there are many reports of nanoscale phase separation in the well-studied $La_{2-x}Sr_xCuO_4$ system,[19,20] including magnetically glassy behavior for a wide doping region.[21,22] However, doping with excess oxygen results in larger phase separated regions with characteristic lengths larger than 100 nm and clear two-phase behavior, which may be attributed to the higher mobility of oxygen atoms and can result in regions larger than 100 nm.[23–25] Again, the larger length scale of the phase separation is associated with the presence of mobile oxygen dopant atoms, which have been shown to take part in ordering transitions of $La_2CuO_{4+\delta}$ down to 200 K.[26] Thus we find that the nature of the dopant ion may play a major role in the varying types of phase separation reported in doped Mott insulator oxides. Another unusual aspect of this system is that we are reporting phase separation between two different FM phases, rather than completely different electronic states. Phase separation between two FM states, while unusual, has been reported for the cobaltite $Pr_{1-x}Ca_xCoO_3$.[27] In that study, two distinct FM regions form at different temperatures, distinguished by short and long range order. The dependence of magnetic ordering on the size of FM clusters underscores the possible role of finite dimension effects in this work.

In summary, we prepared and characterized epitaxial thin films of $SrCoO_x$ with $2.75 \leq x \leq 3.0$, on (1 0 0) and (1 1 1) STO substrates. Electrochemical oxidation produced FM phases similar to those in the bulk, but two simultaneous phases were only observed after several days of deoxidation in atmosphere for thicker (1 1 1) films. The agreement between the measured c lattice constant and that reported for the bulk implies the films are relaxed; meaning the lack of two-phase behavior is not a result of strain. In the bulk materials, the presence of two FM transition temperatures at intermediate oxidation levels implies simultaneous regions of different hole or charge concentrations. Furthermore, the observed phases in both systems correspond to the magnetic phases reported for oxygen contents of x = (3- 1/8 n), with $SrCoO_{3.0}$ (n = 0), $SrCoO_{2.88}$ (n = 1), and $SrCoO_{2.75}$ (n = 2). This suggests the presence of a long range ordering commensurate with the lattice. As the oxygen content in the films is altered, the finite dimensionality normal to the surface may inhibit the formation of an electronically distinct region smaller than 200 nm. In contrast, the thicker (1 1 1) films allow the establishment of large scale, separate regions which phase separate in a similar way as observed in the bulk. This further implies a connection between long range interactions and electronic phase separation in this system.

*Acknowledgements*

This work is supported by the NSF through grant DMR-0907197 (UConn) and A & S Dean's office research funds (UHart).Work at NIU was supported by Great Journey Assistantship (CA) and LnSET NanoScience Distinguished Graduate Fellowships (SAS).

*References*

[1] A. Moreo, S. Yunoki, and E. Dagotto, Science **283**, 2034 (1999).

[2] P.L. Kuhns, M.J.R. Hoch, W.G. Moulton, A.P. Reyes, J. Wu, and C. Leighton, Phys. Rev. Lett. **91**, 95 (2003).

[3] R. Caciuffo, D. Rinaldi, G. Barucca, J. Mira, J. Rivas, M.A. Señarís-Rodríguez, P.G. Radaelli, D. Fiorani, and J.B. Goodenough, Phys. Rev. B **59**, 1068 (1999).

[4] J. Wu, J. Lynn, C. Glinka, J. Burley, H. Zheng, J. Mitchell, and C. Leighton, Phys. Rev. Lett. **94**, 037201 (2005).

[5] C.K. Xie, Y.F. Nie, B.O. Wells, J.I. Budnick, W.A. Hines, and B. Dabrowski, Appl. Phys. Lett. **99**, 052503 (2011).

[6] A. Nemudry, P. Rudolf, and R. Schollhorn, Chem. Mater. **8**, 2232 (1996).

[7] T. Takeda, Y. Yamaguchi, and H. Watanabe, J. Phys. Soc. Jpn. **33**, 970 (1972).

[8] R. Le Toquin, W. Paulus, A. Cousson, and C. Prestipino, J. Am. Chem. Soc. **128**, 13161 (2006).

[9] P. Bezdicka, A. Wattiaux, J.C. Grenier, M. Pouchard, and P. Hagenmuller, Z. Anorg. Allg. Chem **619**, 7 (1993).

[10] Y. Long, Y. Kaneko, S. Ishiwata, Y. Taguchi, and Y. Tokura, J. Phys.: Condens. Matter **23**, 245601 (2011).

[11] N. Ichikawa, M. Iwanowska, M. Kawai, C. Calers, W. Paulus, and Y. Shimakawa, Dalton Trans. **41**, 10507 (2012).

[12] H. Taguchi, M. Shimada, and M. Koizumi, J. Solid State Chem. **29**, 221 (1979).

[13] C. Xie, Ph.D. Thesis, University of Connecticut, Storrs, 2008.

[14] C.K. Xie, J.I. Budnick, B.O. Wells, and J.C. Woicik, Appl. Phys. Lett. **91**, 172509 (2007).

[15] M.A. Torija, M. Sharma, M.R. Fitzsimmons, M. Varela, and C. Leighton, J. Appl. Phys. **104**, 023901 (2008).

[16] J. Mira, J. Rivas, G. Baio, G. Barucca, R. Caciuffo, D. Rinaldi, D. Fiorani, and M.A. Señarís Rodríguez, J. Appl. Phys. **89**, 5606 (2001).

[17] J. Wu and C. Leighton, Phys. Rev. B **67**, 174408 (2003).

[18] S. Kolesnik, B. Dabrowski, J. Mais, M. Majjiga, O. Chmaissem, A. Baszczuk, and J. Jorgensen, Phys. Rev. B **73**, 214440 (2006).

[19] J. Jorgensen, B. Dabrowski, S. Pei, D. Hinks, L. Soderholm, B. Morosin, J. Schirber, E. Venturini, and D. Ginley, Phys. Rev. B **38**, 11337 (1988).

[20] Y.S. Lee, R.J. Birgeneau, M.A. Kastner, Y. Endoh, S. Wakimoto, K. Yamada, R.W. Erwin, and G. Shirane, Phys. Rev. B **60**, 3643 (1999).

[21] F. Chou, N. Belk, M. Kastner, R. Birgeneau, and A. Aharony, Phys. Rev. Lett. **75**, 2204 (1995).

[22] C. Niedermayer, C. Bernhard, T. Blasius, A. Golnik, A. Moodenbaugh, and J.I. Budnick, Phys. Rev. Lett. **80**, 3843 (1998).

[23] H.E. Mohottala, B.O. Wells, J.I. Budnick, W. a Hines, C. Niedermayer, L. Udby, C. Bernhard, A.R. Moodenbaugh, and F.-C. Chou, Nature Materials **5**, 377 (2006).

[24] H. Mohottala, B. Wells, J. Budnick, W. Hines, C. Niedermayer, and F. Chou, Phys. Rev. B **78**, 064504 (2008).

[25] A. Savici, Y. Fudamoto, I. Gat, T. Ito, M. Larkin, Y. Uemura, G. Luke, K. Kojima, Y. Lee, M. Kastner, R. Birgeneau, and K. Yamada, Phys. Rev. B **66**, 014524 (2002).

[26] B.O. Wells, R.J. Birgeneau, F.C. Chou, Y. Endoh, D.C. Johnston, M. a. Kastner, Y.S. Lee, G. Shirane, J.M. Tranquada, and K. Yamada, Zeitschrift Für Physik B Condensed Matter **100**, 535 (1996).

[27] S. El-Khatib, S. Bose, C. He, J. Kuplic, M. Laver, J. Borchers, Q. Huang, J. Lynn, J. Mitchell, and C. Leighton, Phys. Rev. B **82**, 100411 (2010).